# Knowledge Space Framework
An API for representation, persistence and visualization of knowledge spaces

Syed Nasar
*Educational Technology*
*MSCS, Georgia Tech*

*Abstract*—This paper will discuss the challenges in tooling around the management and utilization of knowledge space structures, via standardized APIs for external Adaptive Learning Systems (ALS) to consume. It then describes how these challenges are addressed in a graph based knowledge management framework application designed for external ALSs.

## I. INTRODUCTION

THIS PROJECT IS BASED ON A KNOWLEDGE ONTOLOGY CONCEPT, CALLED KNOWLEDGE SPACES. THE OBJECTIVE OF THIS PAPER IS TO DESCRIBE A KNOWLEDGE SPACE FRAMEWORK BUILT OVER A KNOWLEDGE SPACE PERSISTENCE AND REPRESENTATION GRAPHICAL DATA-STORE.

### A. Concepts

Knowledge space structures are comprised of knowledge states (knowledge items) that the learners learn and explore as they are gaining knowledge. The collection of all the knowledge states capture the organization of the knowledge and is called knowledge structure.

Knowledge state means the set of all problem types (or items) that the learner is capable of solving in ideal circumstances. In other words, a knowledge state is a subset of the domain. The items or problem types used for assessing the learner is called domain of the body of knowledge.

Learning Space is a knowledge structure if it satisfies the two following conditions, learning smoothness and learning consistency.

The idea of a learning path is that every student progresses across some learning path. With the passing of time the learner successively masters the items encountered along the learning path.

### B. Representing Knowledge Space Structures

A knowledge map can be expressed as a directed graph in which the nodes are concepts and the edges specify prerequisite relationships between the concepts. A Knowledge map primarily serves as a reference tool which typically has associated content with each concept, e.g. a video tutorial or a lesson, which a student would use to learn the material.

A common pattern was recognized in how various knowledge structures can be represented. It was noticed that there are mainly 3 important elements that can represent most of the knowledge structures. These are:
- *Concept*
- *Relationship between concepts*
- *Associated content(s) with a Concept or Relationship*

A graph is made up of the following 3 main components:
- *Node*
- *Edge*
- *Labels associated with a Node or an Edge*

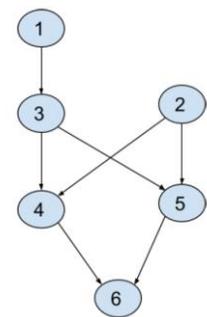

When compare with the knowledge structures and the structure of a graph, there appears a strong similarity. It was realized that knowledge structures can be represented using graph structures as described below (*refer Diagram depicting a graph structure*):
- Nodes can represent Concepts
- Edges can specify Relationships
- Associated content with each Concept or Relation can be added as labels to each Node/Edge respectively.

### C. Knowledge Space Framework Overview

This paper describes a Knowledge Space Framework (KSF) system. KSF exposes the underlying graphical knowledge structures via web based APIs (Restful). The KSF APIs allow the consuming systems to create, visualize, and navigate the entire graphical representation of the knowledge space structures.

The diagram below shows the components of KSF and its underlying graphical store framework:

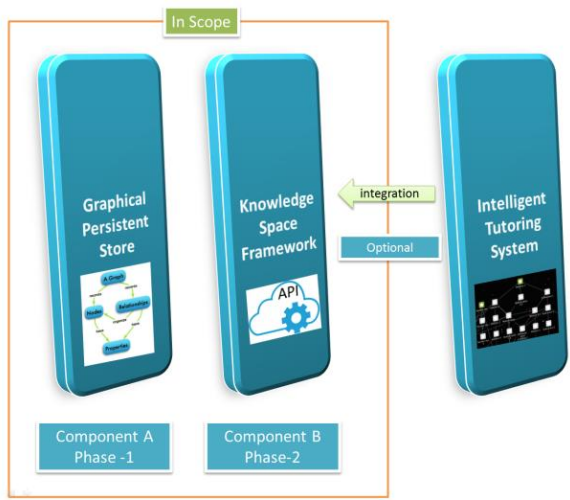

Diagram A: The conceptual view of the various components developed as part of the KSF

## II. MOTIVATION

### A. Inspiration

In 1985, Doignon and Falmagne[1] came up with the idea of knowledge space theory (KST) to overcome the drawbacks of the psychometric[2] method. They mention the foundational idea in their paper which explains "*The basic idea is that an assessment in a scholarly subject should uncover the individual's 'knowledge state', that is, the exact set of concepts mastered by the individual.*". So the main motivation for this project came from the idea of how to represent, persist and visualize the knowledge space structure elements.

### B. Challenge of Representation, Persistence and Visualization of knowledge spaces

A difficult problem here is to physically implement the various constructs of knowledge spaces and learning spaces. The challenge is to implement this in a manner that it is easy to manage and interact with, easy to search and present visually the underlying information (without the overhead of writing complex abstraction layers); can be scaled and distributed across multiple machines, should be highly performant, and most importantly should be able to represent most of the concepts that make up a knowledge space.

### C. Challenge of Abstraction and Navigation of knowledge spaces

Another difficulty is about how to abstract these constructs for external systems to navigate and interact with them. The challenge is to make the navigation as seamless and as intuitive as possible. Currently, there is also a lack of any open framework available that can be built on top of a data store which can be leveraged by adaptive learning systems.

### D. Ideation

At the exploration stages of this initiative, it was recognized that there is a need for a standardized tool that would provide a framework which would abstract out the complexities of knowledge space constructs and allow different types of ALSs, e.g. ITSs etc, to interact with it for managing, administering and navigating the knowledge spaces. Thus the key idea was to develop such a knowledge space framework.

## III. RELATED RESEARCH AND TOOLING

This section will discuss some of the work that were explored to understand what related research and tooling are available in this area, specific to the problem defined earlier.

### A. Related research and tooling

There are some intelligent tutoring systems that have developed similar knowledge representations. Some examples are Khanacademy[3], Metacademy[4] and ALEKS[5]. Many of these implementations are developed using generic tools, which does not attempt to align the technology with the solution. The data-store layers for these ITSs were part of an Intelligent Learning System, and do not have the ability to be reused for consumption or integration with other ITSs.

No related tooling could be found that were aimed specifically at providing a reusable underlying knowledge space management framework. Upon exploring the various players in the market it appears that many of the tools were homegrown tools developed using generic technologies e.g. web frameworks, javascript libraries, GUI based graphing tools etc, but there were no technologies that aligns naturally to a graph based structure.

The goal was to find a data-store technology which minimizes any impedance mismatch[6] between the knowledge

---

[1] Doignon, J.P.; Falmagne, J.Cl. (1985), "Spaces for the assessment of knowledge", International Journal of Man Machine Studies

[2] Psychometric method is a technique to accomplish competency based assessments. Within this method standardized test results are used to place a learner in one of the ordered categories out of many.

[3] Khan Academy, Assessing student mastery, retrieved from - https://www.khanacademy.org/newandnoteworthy/v/overviewofkhanacademyorg

[4] Metacademy, retrieved from - https://metacademy.org/about

[5] ALEKS, retrieved from - https://www.aleks.com/about_aleks/overview

[6] Impedence Mismatch - The object-relational impedance mismatch is a set of conceptual and technical difficulties that are encountered when a

spaces and how the database technology represents such structures. In other words, there is a need for a tool that stores and represents knowledge space structure constructs in a graphical structure.

*B. Abstraction and Navigation of knowledge spaces related work*

There is very sparse work in this area. No tooling could be found having the ability of providing a standardized API for managing, administering, navigating and visualizing a knowledge space repository. Most solutions, mentioned above, are tightly coupled with the tutoring applications they were built for.

IV. PLAN AND IMPLEMENTATION

*A. Planned Solution*

The scope was to develop a knowledge space persistence and representation data-store and develop an open framework API on top of it, which should be flexible and can represent an expressive ontology, and should allow the representation of a diverse set of knowledge space constructs which can be easily related to each other.

*B. High level design of the components*

This section discusses the high level design. The system is comprised of two main subsystem components:

Graphical Persistent Store (abbreviated as GPS) – This is a graphical persistent data-store that represents and persists knowledge structures comprised of knowledge states. The component also allows the visualization of the hierarchical structure of the knowledge states, and how they are related, using the labeled nodes and edges of the graphs.

Knowledge Space Framework (KSF) and Restful API – This component is a reusable knowledge space management framework that can be leveraged by ALSs using standardized APIs. It has a standardized Restful API which allows creation, navigation and querying of the knowledge structures.

A web-based ITS prototype was also developed additionally to showcase how the KSF APIs can be integrated with an external ALS. The ITS component is entirely developed as a client side (browser based) web application.

---

RDBMS is being used by a program written in an object-oriented programming language or style, particularly when objects or class definitions are mapped in a straightforward way to database tables or relational schemata. Definition retrieved from - https://en.wikipedia.org/wiki/Objectrelational_impedance_mismatch

*C. What was planned?*

The two components were planned to be developed in distinct phases:
  a. PHASE-I – This phase was aimed at developing the Graphical Persistent Store (GPS).
  b. PHASE-II – Develop the Knowledge Space Framework (KSP) which would allow external ALSs to interact with the GPS component.
  c. Optional PHASE – Develop a prototype to showcase how an external ALS, a skeletal version of an ITS, would interact with the Knowledge Space Framework.

*D. Implementation*

All three components were developed within a span of 8 weeks, between March, 2016 and April, 2016. The implementation also includes the additional optional integration component that showcases how the KSF can be integrated with an external ITS.

*E. Architecture*

This section discusses how each subsystem has been designed and developed. The following diagram shows the two main components:

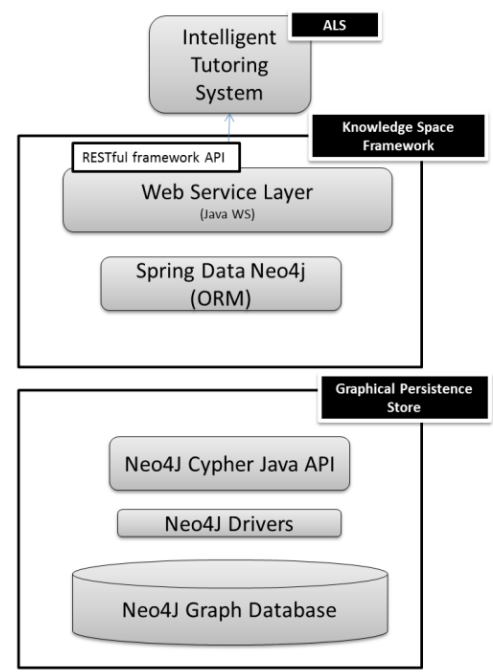

The GPS (graphical persistence store) has a graphical database, called Neo4j, at its core. This database is a server based graph database, which is accessed by using the drivers provided by the Neo4j SDK[7]. The technologies used were the Neo4j Cypher Java API to access the Neo4j drivers, which would in turn provide direct data manipulation abilities to the client program.

---

[7] SDK – software development kit, a form of programming interfaces for external systems to integrate with.

The KSF component is developed on top of the GPS component. KSF is developed using an ORM (Object-relational mapping) library called, Spring Data Neo4j, which would interact with the GPS layer's API. The KSF exposes the graph manipulation and navigation interfaces as Restful APIs using Spring and Jackson with D3 libraries.

The ITS component is developed using JQuery and d3.js.

## V. RESULTS AND ACCOMPLISHMENTS

### A. Accomplishments

As a final product, two main components were developed, GPS and KFS. In addition a toy prototype was built to mimic an ITS client to validate the Restful APIs of the knowledge space constructs exposed by KSF.

The GPS component, as mentioned earlier, is a graphical database at its core. The component represents the following main knowledge space constructs as of today (these can easily be extended further to incorporate additional knowledge space concepts) -

- Nodes:
    - Knowledge State (attribs: *title, topic, released, tags*)
    - Learner (attribs: *name, reviewed-on*)
        - Student
        - Instructor
    - Lesson or Challenge
        - Associated to knowledge states and linked to students as 'reviewed' and to instructors as 'developed'.
- Relations (Edges of the nodes):
    - Reviewed (attribs: *lessons*)
    - Precedes (attribs: *weight*)

The diagram below shows how the various knowledge constructs are represented in the graphical store:

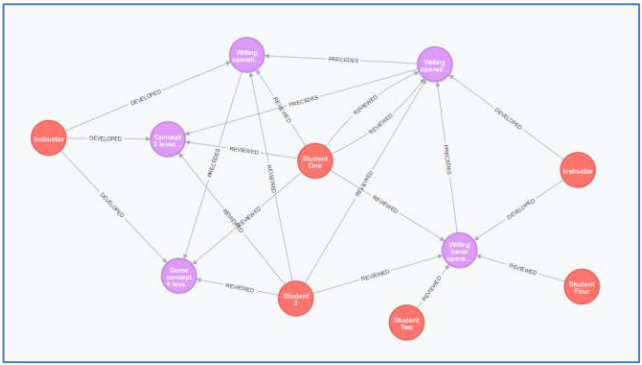

The GPS subsystem is a multilayered system. It exposes the graphical data-store using Java with Neo4j Cypher Java APIs. This library is utilized in the repository layer of the software.

There are two main layers in the KSF subsystem. The lower layer is developed using the Spring Data Neo4j ORM. This layer invokes the GSF Cypher APIs. Here is a representative code showing how this layer interacts with the Cypher API:

```
@Query("MATCH (m:KnowState) WHERE m.topic =~ ('(?i).*'+{topic}+'.*') RETURN m")
Collection<KnowState> findByTopicContaining(@Param("topic") String topic);
```

The topmost layer, which lies at the boundary of the KSF is developed using Spring and Jackson with D3 libraries. This exposes the underlying services as Restful web services. Here are some of the representative operations that are exposed via this layer:

| API Method | Description |
| --- | --- |
| KnowStateRepository .findByTopicContaining | Search knowledge states by topic |
| LearnerRepository. findByName | Retrieve learners by name |

The following screenshot shows a sample output of the KSF API, when searching for Knowledge States:

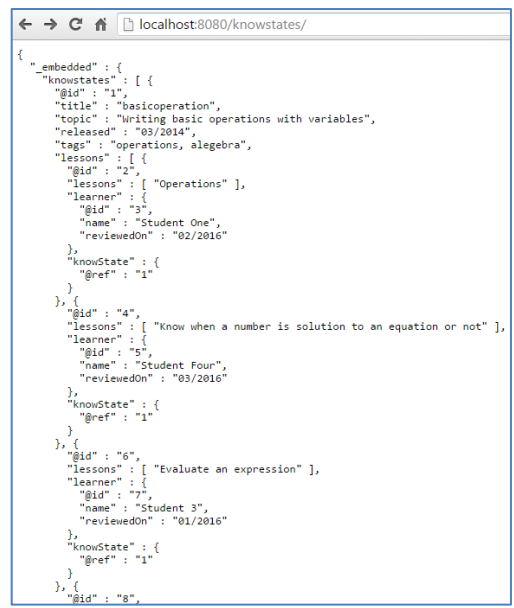

The web-based ITS prototype is built using javascript libraries, JQuery and d3.js. This component directly queries the Resful APIs of KSF system. Here is a representative screenshot of a search scenario from this ITS prototype:

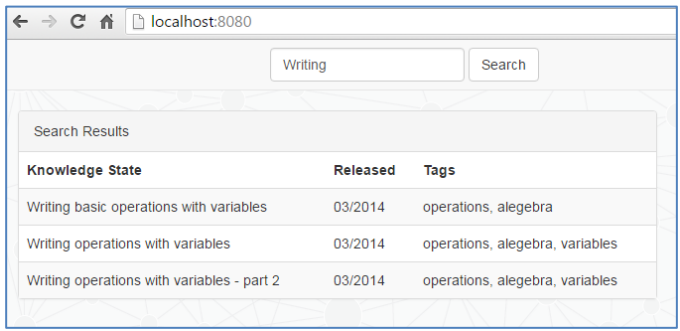

A sample screenshot of the complete web page:

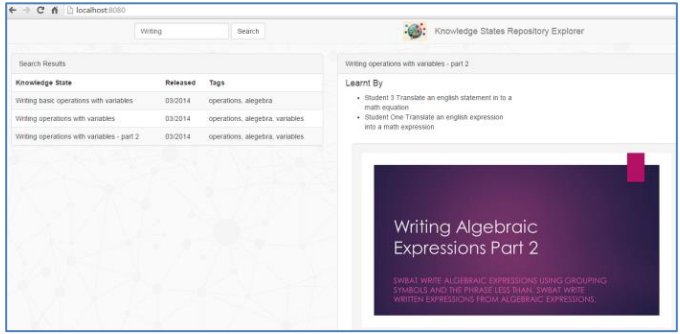

*Note: This component is not part of the main framework/project. This is developed only to simulate a sample client application's interaction with KSF.*

B. *What can be developed further*

The advantage of developing this framework on top of a graphical database is to mainly provide the ability to extend the framework to accommodate concepts that are not part of the current implementation. The nodes and edges can be associated with new concepts, represented by new nodes and edges. Additional attributes can be accommodated by adding to existing nodes and edges, to further enrich these concepts.

The flexible Restful API allows any client to interact with this framework, as long as it can integrate with Restful interfaces.

Some of the concepts, especially around Bayesian knowledge tracing can be incorporated by adding additional probability scores to the edges of learning.

If multiple ALSs can adopt this framework, then a knowledge space information exchange can be developed, which would allow sharing of learning metrics across multiple learning systems. These can further be leveraged for applying machine learning against the learner population data.

## VI. APPENDIX

### ACKNOWLEDGEMENT

This paper acknowledges the invaluable support and guidance provided by Ken Brooks, the mentor for this project. His intuitive and creative ideas helped the ideation of the proposal and led to a valuable and well-rounded product.